\documentclass[aip,apl,superscriptaddress,amsmath,amssymb,reprint]{revtex4-1}
\usepackage{graphicx}
\usepackage{dcolumn}
\usepackage{bm}
\usepackage{textcomp}
\usepackage[utf8]{inputenc}

\begin{document}

\title{Versatile module for experiments with focussing neutron guides}

\author{T. Adams}
\address{Physik-Department,
  Technische Universit\"at M\"unchen, 85748 Garching, Germany}
  
\author{G. Brandl}
\address{Physik-Department,
  Technische Universit\"at M\"unchen, 85748 Garching, Germany}
\address{Heinz Maier-Leibnitz Zentrum, FRM II, 
  Technische Universit\"at M\"unchen, 85748 Garching, Germany} 

\author{A. Chacon}
\address{Physik-Department,
  Technische Universit\"at M\"unchen, 85748 Garching, Germany}
\address{Heinz Maier-Leibnitz Zentrum, FRM II, 
  Technische Universit\"at M\"unchen, 85748 Garching, Germany} 

\author{J. N. Wagner}
\address{Physik-Department,
  Technische Universit\"at M\"unchen, 85748 Garching, Germany}
\address{Heinz Maier-Leibnitz Zentrum, FRM II, 
  Technische Universit\"at M\"unchen, 85748 Garching, Germany} 

\author{M. Rahn}
\address{Physik-Department,
  Technische Universit\"at M\"unchen, 85748 Garching, Germany}
\address{Heinz Maier-Leibnitz Zentrum, FRM II, 
  Technische Universit\"at M\"unchen, 85748 Garching, Germany} 

\author{S. M\"uhlbauer}
\address{Physik-Department,
  Technische Universit\"at M\"unchen, 85748 Garching, Germany}
\address{Heinz Maier-Leibnitz Zentrum, FRM II, 
  Technische Universit\"at M\"unchen, 85748 Garching, Germany} 

\author{R. Georgii}
\address{Physik-Department,
  Technische Universit\"at M\"unchen, 85748 Garching, Germany}
\address{Heinz Maier-Leibnitz Zentrum, FRM II, 
  Technische Universit\"at M\"unchen, 85748 Garching, Germany} 

\author{C. Pfleiderer}
\address{Physik-Department,
  Technische Universit\"at M\"unchen, 85748 Garching, Germany}

\author{P. B\"oni}
\address{Physik-Department,
  Technische Universit\"at M\"unchen, 85748 Garching, Germany}

\date{\today}

\begin{abstract}
We report the development of a versatile module that permits fast and reliable use of focussing neutron guides under varying scattering angles. A simple procedure for setting up the module and neutron guides is illustrated by typical intensity patterns to highlight operational aspects as well as typical parasitic artefacts. Combining a high-precision alignment table with separate housings for the neutron guides on kinematic mounts, the change-over between neutron guides with different focussing characteristics requires no readjustments of the experimental set-up. Exploiting substantial gain factors, we demonstrate the performance of this versatile neutron scattering module in a study of the effects of uniaxial stress on the domain populations in the transverse spin density wave phase of single crystal Cr.
\end{abstract}
\maketitle

Neutron scattering allows to solve some of the most pressing scientific challenges, when studies of tiny sample dimensions, e.g., as constraint by the sample environment or sample quality, are made possible. An elegant technical solution for such neutron scattering studies of tiny samples are elliptically or parabolically shaped neutron guides, which focus the neutron beam on a small spot\cite{Hils:2004,Kardjilov:2005}. As an essential added advantage this minimises the background signal at the same time. Extensive numerical studies and proof-of-principle experiments have demonstrated the potential power of this method\cite{2014:Boeni:JPCS}. However, the use of focussing neutron guides on a routine basis has so far been hampered severely by parasitic signal contributions when the rather stringent requirements of alignment cannot be satisfied. Because neutron scattering experiments are performed at large-scale facilities on typical time scales of several days, the perhaps most important practical aspect when using focussing neutron guides concerns the ability to set them up reliably in the shortest possible time.

An accurate and reliable alignment is required due to the combination of the small dimensions of the spot size (well below ${\rm mm^2}$) and the size of the focussing neutron guides (overall length $\sim0.5\,{\rm m}$; cross-section: $\sim {\rm cm^2}$) as compared to the large size of the neutron scattering instruments, which typically exceeds several meters. The neutron guides must thereby stay aligned with respect to the beam axis to better than $0.1^{\circ}$ for all scattering angles, while maintaining the distance of the neutron guides to the sample and the optical axis of the neutron scattering instrument to better than a fraction of the spot size, i.e. better than $\simeq 0.1 {\rm\,mm}$. 

In this Letter we report the development of a module for experiments with focussing neutron guides that satisfies these criteria. The module we have developed may be set up in as short a time as a few hours on completely different neutron scattering instruments. Once the module has been set up, it permits changes between different focussing guides without loss of alignment. This allows to adapt the spot size and focal length instantly. To demonstrate the improvements in measurements of small samples we have studied the uniaxial pressure dependence of the domain populations in the transverse spin density wave state of single-crystal Cr, where we achieved an increase of the intensity by a factor of four.

\begin{figure}
\includegraphics[width=0.95\linewidth]{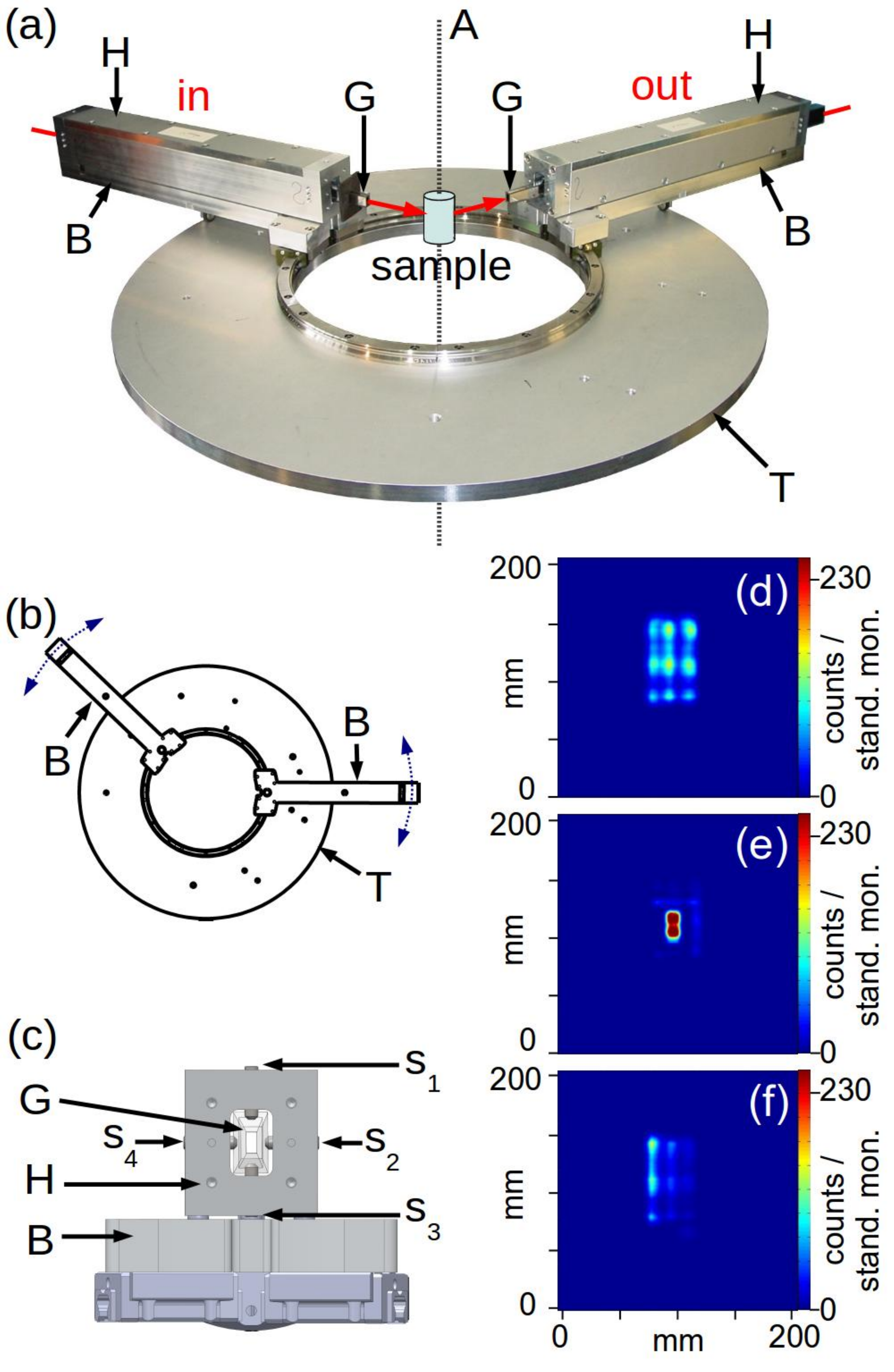}
\caption{\label{figure1}
Versatile module for experiments with focussing neutron guides. (a) Photograph of the alignment table (T) and the base plates (B), which support the housings (H) with the focussing neutron guides (G) using kinematic mounts. The sample is centered at the axis (A). (b) Top view of the alignment table (T) and the two rotatable base plates (B). (c) Base plate (B), housing (H), and focussing neutron guide (G) as seen from the sample. The position of the neutron guides may be adjusted by set screws (S$_i$) (for clarity only four of these screws are marked). (d) Intensity pattern of the direct neutron beam as measured 1\,m behind the sample using one adjusted focussing neutron guide for the incident beam. The rectangular distribution of nine maxima illustrates the phase space distribution of the neutrons at the location of the sample. (e) Intensity pattern of the direct neutron beam with two adjusted focussing guides. At the exit of the second guide the phase space distribution is homogeneous. (f) Asymmetric intensity distribution of the direct neutron beam for one misaligned guide (cf panel (d)), reflecting an inhomogeneous phase space as compared with panel (d). 
}
\end{figure}

Shown in Fig.\,\ref{figure1}\,(a) are the main components of the module. A high-precision alignment table (T), providing a circular slide rail, supports two freely-movable base plates (B). Housings (H) containing the focussing guides (G) can be reproducibly installed on these plates (B) using kinematic mounts.  While the alignment table (T) and the base plates (B) link the module to the neutron scattering instrument and control the scattering angles as depicted schematically in Fig.\,\ref{figure1}\,(b), the neutron guides are aligned and attached to the housings (H) by set-screws (S1) through (S4). This allows for a highly reproducible alignment and fast turn-around times, because, (i), the combination of the alignment table (T) and base plates (B) provides a rigid and accurate support structure, (ii), the focussing neutron guides need to be aligned only once with respect to the housings (H), and, (iii) the positioning of the housings (H) by virtue of the kinematic mounts is extremely accurate and reproducible (better than $0.01\,{\rm mm}$ and $0.01^\circ$).

To reduce the time for setting up the module on the neutron scattering instrument (representing an important operational cost factor), the focussing neutron guides may be aligned with respect to the housings by optical methods. For this purpose an accurately machined plate with a  pin on its central axis is bolted to the alignment table (T). A bespoke optical system comprising of a laser and a set of lenses illuminates  the neutron guide (G) uniformly at the back-end of the housing (H). The position of the neutron guide (G) may then be adjusted by the set-screws (${\rm S_i}$) until, (i) the focal point of the laser light coincides with the tip of the pin located at the sample position, and, (ii) the image of the laser beam at a location well behind the sample is symmetric. The preparation of neutron guides with different focussing properties, each with their own housing, facilitates instant changes between different focussing characteristics during the experiments. The aligned module is, finally, mounted on the sample table of the neutron scattering instrument such that its centre coincides with the centre of the sample table, i.e., the rotation axis of the sample goniometer to within $\simeq 0.1$\,mm. This procedure proves to be highly reproducible and very fast.

\begin{figure}
\includegraphics[width=0.95\linewidth]{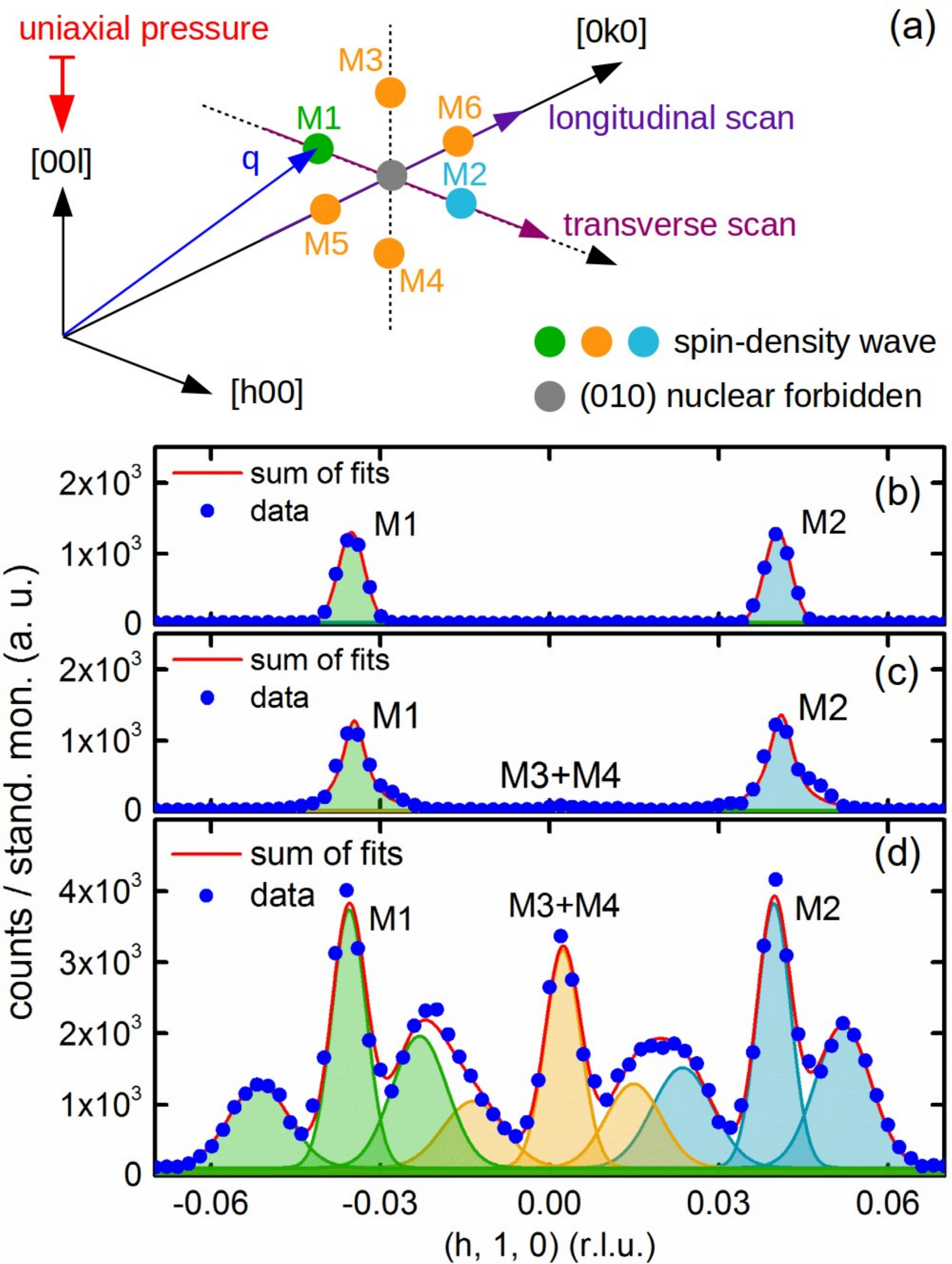}
\caption{\label{figure2}
Typical momentum-scans in Cr using focussing neutron guides. All data are shown as measured without background subtraction. (a) Schematic depiction of the magnetic Bragg peaks in the SDW state (blue, yellow and green) in the vicinity of the forbidden $(010)$ nuclear Bragg peak and the direction of transverse and longitudinal momentum scans. (b) Transverse $q$-scan without neutron guides using a counting tube. (c) Transverse $q$-scan with one neutron guide using a counting tube. (d) Transverse $q$-scan with two neutron guides using a counting tube. Under changing scattering angle the second neutron guide captures changing portions of the phase space distribution of the neutron beam at the location of the sample (cf Fig.\,\ref{figure1}\,(d)). This results in a main maximum with side maxima. 
}
\end{figure}

Typical improvements as well as parasitic effects when using focussing neutron guides may be illustrated with data recorded at the diffractometer MIRA at FRM II. Neutrons with a wavelength $\lambda=4.5\,{\rm \AA}\pm 2\%$ were recorded with an area detector. The neutron guide elements used for this test were part of a 2\,m long elliptic guide with a critical angle of reflection, $m = 3$ times the critical angle of Ni. Each guide element was 500\,mm long with a focal length of $80\,{\rm mm}$ and a focal spot size of $\sim1\,{\rm mm^2}$. The rectangular cross sections at the front and and back end of the guides were $9\times 18\,{\rm mm^2}$ and $4\times 8\,{\rm mm^2}$ (width$\times$height), respectively \cite{2006:Muehlbauer:PhysicaB}.

Fig.\,\ref{figure1}\,(d) shows the intensity distribution when the detector is placed 1\,m behind one well-aligned neutron guide, which focusses the incident beam on the sample position. The detector image shows that the phase space assumed by the neutron beam, which is homogeneous at the entry of the guide, becomes inhomogeneous at the location of the sample, consisting of a rectangular pattern of nine maxima. The central maximum of this distribution is dominated by the direct beam, whereas the remaining eight maxima may be attributed to the sides and the corners of the guide \cite{2014:Boeni:JPCS}. It is important to note that a second guide with its optical axis parallel to the first guide and with its focal point at the sample position, transforms the inhomogeneous phase space of the neutron beam at the sample location such that it is homogeneous again at the exit of the guides and thus at the detector as shown in Fig.\,\ref{figure1}\,(e).

\begin{figure}
\includegraphics[width=0.95\linewidth]{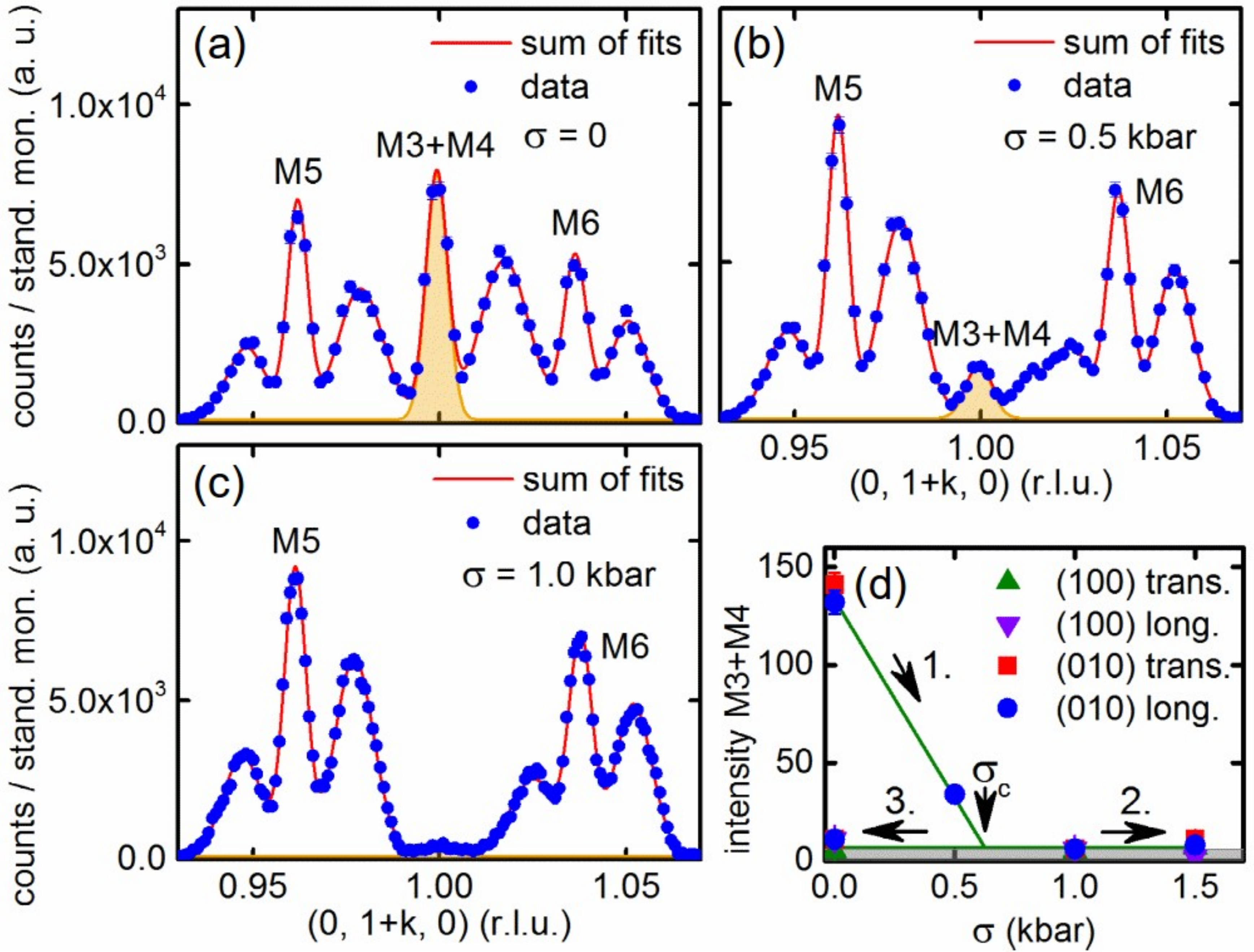}
\caption{\label{figure3}
Effect of uniaxial pressure on the domain distribution in the transverse SDW of Cr as observed in longitudinal $q$ scans at ambient temperature. With increasing uniaxial pressure the domain population in the direction of the uniaxial stress axis (M3 and M4) is depopulated and vanishes irreversibly above a critical pressure $\sigma_c\approx0.6\,{\rm kbar}$. All data are shown as measured without background subtraction.
}
\end{figure}

When the optical axis of the focusing guide and the incident beam are \textit{not parallel}, representing the most important example of a misalignment, the inhomogeneous phase space at the location of the sample is no longer symmetric as shown in Fig.\,\ref{figure1}\,(f) (here the same detector set-up was used as in Fig.\,\ref{figure1}\,(d)). In turn, this may lead to an erroneous determination of the intensity of the diffraction peaks and the scattering angles. A discussion of this artefact for linearly tapered neutron guides has been reported in Ref.\,\cite{Boeni:1998}.

The uniaxial pressure dependence of the domain populations in the transverse spin density wave (SDW) in Cr illustrate the improvements in beam intensity. At the N{\'e}el temperature, $T_{\rm N} = 311\,{\rm K}$, Cr undergoes a weak first order transition from paramagnetism to SDW order\cite{1988:Fawcett:RevModPhys}, which is characterised by incommensurate wave vectors $Q^{\pm} = (0, 0, 1 \pm\xi)$ with $\xi \approx 0.046$. It has long been predicted, that the transition at $T_{\rm N}$ changes from first to second order under tiny uniaxial pressures, where $T_{\rm N}$ remains essentially unchanged \cite{1982:Barak:JPhysF}. The uniaxial pressure dependence thereby serves to probe the nature of the first order behavior, which is suggestive of issues related to fluctuation-induced first order in chiral helimagnets\cite{2013:Janoschek:PRB,2014:Bauer:PRL}. However, a first neutron scattering experiment was inconclusive \cite{1984:Fawcett:JPhysF}.

We used a He-activated bellow system \cite{Pfleiderer:RSI1996,Waffenschmidt:PRL1999,Chacon:Thesis} to apply uniaxial pressure, $\sigma$, along the $[001]$ axis of a small cubic sample ($2\times2\times2\,{\rm mm^3}$). Our study was performed at MIRA (FRM II) using a counting tube instead of the area detector. Data of the magnetic Bragg peaks were collected in the vicinity of the forbidden $(010)$ nuclear Bragg spot, as schematically shown in Fig.\,\ref{figure2}\,(a). Higher order scattering was suppressed with a Be filter. For what follows, it is important to appreciate that the changes of intensity are not due to a change of $T_{\rm N}$, which is essentially unaffected. We also note, that all data are shown as measured without background subtraction.

A typical transverse scan at ambient temperature without neutron guides is shown in Fig.\,\ref{figure2}\,(b), where the magnetic satellites at M1 and M2 are seen. The same scan as recorded with one well-aligned neutron guide in front of the sample is shown in Fig.\,\ref{figure2}\,(c). Because of the inhomogeneous phase space distribution of the neutrons at the sample (cf. Fig.\,\ref{figure1}\,(d)) and the small diameter of the counting tube the absolute value of the intensity appears essentially unchanged with faint additional shoulders and very weak intensity attributed to M3 plus M4. 

The nature of the faint shoulders in Fig.\,\ref{figure2}\,(c) becomes evident when the same $q$-scan is recorded with two well-aligned neutron guides as shown in Fig.\,\ref{figure2}\,(d). Here a large increase of intensity is observed. As an important additional feature each magnetic Bragg peak (M1 and M2) consists now of an intensity maximum and two side peaks, because the second neutron guide captures different parts of the inhomogeneous phase space assumed by the neutron beam at the sample location as the scattering angle changes. A detailed analysis reveals that these side-peaks are rather generic with a ratio of the intensities of the main peak to the side peaks of three to one. Moreover, the tails of the peaks at M3 plus M4 significantly gain intensity, because the inhomogeneous phase space distribution at the sample location also yields a larger effective divergence. 

To track changes of the domain populations due to uniaxial pressure directly we performed longitudinal scans as shown in Fig.\,\ref{figure3}, where the pairs M3/M4 and M5/M6 reflect the behaviour parallel and perpendicular to the pressure axis, respectively. These experiments were performed at ambient temperature. As shown in Fig.\,\ref{figure3}\,(a) the intensity for each magnetic spot gives again rise to a main peak and two side peaks. With increasing pressure, the contribution due to M3 and M4 decreases and vanishes above a critical pressure $\sigma_{\rm c}\approx0.6\,{\rm kbar}$ (step 1 followed by step 2 in Fig.\,\ref{figure3}\,(d)), while that due to M5 and M6 increases $\sim30\,\%$ (Fig.\,\ref{figure3}\,(b) and (c)). We have confirmed this behaviour in additional transverse scans and longitudinal and transverse scans after rotating the crystal by $90^{\circ}$. 

When decreasing the pressure again to zero at unchanged temperature (step 3 in Fig.\,\ref{figure3}\,(d)), the domains M3/M4 remain unpopulated. This is consistent with a uniaxial pressure-induced symmetry breaking as predicted theoretically. In turn, with the reduction of the number of degenerate domain populations the phase space occupied by critical fluctuations also decreases such that the effects of interactions between the fluctuations becomes smaller. This is expected to permit the transition to become second order. However, while the suppression of domain populations are consistent with theory \cite{1982:Barak:JPhysF} a detailed search for uniaxial stress induced criticality and the predicted triciritical point will require further work planned for the future. These studies will be also required to clarify the effects of stress inhomogeneities across the sample as well as direct evidence of the dynamical properties of the spin density. 

In our experiments we have readily achieved gains of intensity of a factor of four as evident from the peak intensities shown in Fig.\,\ref{figure2}. For these estimates we consider the signal as measured, since the background was already very low. The full benefit of using focussing neutron guides will unfold in the presence of large background contributions due to the sample environment. This concerns, for instance, much reduced sample dimensions or high pressure experiments. Namely, for Cu:Be clamp cells \cite{Pfleiderer:JPCM} typically 80\,\% of the neutrons are absorbed or scattered, underscoring the need for very efficient background reduction. 

The biggest benefit of using focussing neutron guides may finally be expected, when the artefacts arising from the combination of inhomogeneous phase space and increased beam divergence at the sample location reported here are not important. A prominent example are inelastic neutron scattering studies, where large beam divergences are favourable. Here major improvements of the signal to noise ratio exceeding well over an order of magnitude are expected. Considering the advantages of signal gain and background reduction demonstrated in this Letter, the versatile module reported here will significantly broaden the scope of such studies.

We gratefully acknowledge support by R. Schwikowski, A. Mantwill, M. Wipp and the team of FRM~II.  Financial support through the collaborative research network TRR80 of the German Science Foundation (DFG) is gratefully acknowledged. TA, GB and AC acknowledge financial support through the TUM graduate school. 


\end{document}